# Prediction of Drug-Induced TdP Risks Using Machine Learning and Rabbit Ventricular Wedge Assay

Nan Miles Xi[1], Dalong Patrick Huang[2]

[1] *Department of Mathematics and Statistics, Loyola University Chicago, Chicago, IL 60660 (e-mail: mxi1@luc.edu)*
[2] *Office of Biostatistics, Office of Translational Science, Center for Drug Evaluation and Research, US Food and Drug Administration, Silver Spring, MD 20993 (email: Dalong.Huang@fda.hhs.gov)*

**ABSTRACT**

The evaluation of drug-induced Torsades de pointes (TdP) risks is crucial in drug safety assessment. In this study, we discuss machine learning approaches in the prediction of drug-induced TdP risks using preclinical data. Specifically, the random forest model was trained on the dataset generated by the rabbit ventricular wedge assay. The model prediction performance was measured on 28 drugs from the Comprehensive In Vitro Proarrhythmia Assay initiative. Leave-one-drug-out cross-validation provided an unbiased estimation of model performance. Stratified bootstrap revealed the uncertainty in the asymptotic model prediction. Our study validated the utility of machine learning approaches in predicting drug-induced TdP risks from preclinical data. Our methods can be extended to other preclinical protocols and serve as a supplementary evaluation in drug safety assessment.

# 1. INTRODUCTION

Torsades de pointes (TdP) is a potentially fatal ventricular arrhythmia that can be induced as a side effect by cardiovascular and noncardiovascular drugs (Stockbridge et al, 2013; Blinova et al, 2018; Liu et al, 2021). The identification of TdP risk is a crucial step in the assessment of new drugs before they reach the market (Stockbridge et al, 2013). Several preclinical paradigms have been promoted to identify the drug-induced TdP risk *in vitro*. Among them, the rabbit ventricular wedge assay (RVWA) is one of the cutting-edge methods (Liu et al, 2006; Sager et al, 2014; Fermini et al, 2016). It uses isolated arterially perfused rabbit ventricular wedge preparation as the proarrhythmic model and measures the electrophysical signals resulting from drug effects at the cellular level (Liu et al, 2021) Although efforts have been largely invested in the experimental development of RVWA, a comprehensive machine learning framework that provides an accurate prediction of drug-induced TdP risks from experimental observations is still lacking. Previous work either relied on nonstatistical methods based on a single predictor or logistic regression to predict TdP risks (Liu et al, 2013; Li et al, 2019). Such methods failed to utilize the multiple predictors generated from experiments or ignored the nonlinear relationship between predictors and TdP risks.

In this study, we proposed a random forest model to predict drug-induced TdP risks from RVWA data with high accuracy. The model performance on new drugs was evaluated by leave-one-drug-out cross-validation (LODO-CV). In addition to the point estimate, the uncertainty of model performance was quantified by stratified bootstrap. We also examined the model interpretation through permutation predictor importance. A sensitivity analysis was conducted to investigate the impact of potential outlier drugs on model performance. We performed control analysis by selecting one high-risk drug and one low-risk drug or placebo as positive and negative controls, thereby providing further model validation.

The proposed modeling strategies were evaluated on an isolated arterially perfused rabbit ventricular wedge preparation dataset. The dataset contains the electrophysiological signals of 28 drugs from the Comprehensive In Vitro Proarrhythmia Assay (CiPA) initiative with low or high TdP risks (Supplementary Table S1). The random forest model achieved the prediction accuracy of 0.8750 by observations and 0.9286 by drugs. Our work is one of the first attempts to build multivariate statistical models with interpretability and uncertainty measurements and can predict the drug-induced TdP risk accurately from *in vitro* experimental data. The proposed modeling and evaluation methods can be easily extended to new datasets and experimental protocols. The result of model prediction could be used as supplementary evidence in the assessment of drug safety.

## 2. DATASETS

The dataset used in this study was generated from a blinded validation of RVWA for the assessment of drug-induced proarrhythmia (Liu et al, 2021). In this experiment, 28 drugs with known low, intermediate, or high TdP risk were applied to RVWA. Electrophysiological responses to 28 drugs at different concentrations were recorded, generating 16 predictors in the dataset ([Supplementary Table S2](#)). For each drug, four observations were collected in one laboratory under the same biological model and electrophysiological platform. There are 112 observations in this dataset (four observations × 28 drugs), and we will refer to it as the wedge dataset in the following text.

## 3. METHODS

### 3.1 Data Preprocessing

We first merged the intermediate and high risks to a new high risk category because the definition of intermediate risk is relatively arbitrary compared with low and high risk. In addition, previous studies have shown that the classification between intermediate and high risk is a major source of error in the predictive model (Xi et al, 2021). Merging these two categories will increase the model performance without missing the main goal of identifying drugs with a certain degree of TdP risks (i.e., the original intermediate and high risks). Second, we utilized the association among predictors to impute missing values in the wedge dataset. A bagging model (Breiman, 1996) was trained by treating predictors with missing values as dependent variables and other predictors as independent variables. The missing values then were imputed by the prediction result of bagging models. It is worth noting that in the imputation process, we did not use the risk information of 28 drugs. Finally, to reduce the redundant predictors in the wedge dataset, we calculated the correlation between each predictor pair and identified the predictor pairs with a Pearson correlation coefficient greater than 0.9. Between the two predictors in such pairs, we dropped one that has a larger mean absolute correlation with other predictors. After the reduction of redundant predictors, there are six predictors left in the wedge dataset.

### 3.2 Model Evaluation Strategy

The model evaluation aims to estimate the prediction accuracy of any new drug outside the current wedge dataset. A valid evaluation strategy should generate an unbiased estimation of the model performance, irrespective of the exact model form. To achieve this goal, we designed a leave-one-drug-out cross-validation (LODO-CV) described in [Figure 1](#) (Hastie, Tibshirani, and Friedman, 2009). Specifically, we trained a predictive model on observations of 27 drugs and predicted the risk of observations in the one left-out drug. We repeated the previous process by predicting each drug with models trained on the other 27 drugs. The final model prediction accuracy was calculated by combining the prediction result of all 28 drugs.

Mathematically, let $N$ be the number of drugs in the dataset ($N$ = 28 in the wedge dataset) and $M$ the number of observations per drug ($M$ = 4 in the wedge dataset). Denote $\hat{f}^{-k}$ as the model trained on the dataset with the $k$th drug removed. Let $(x_m^k, y_m^k)$ be the $m$th observation of drug $k$, where $x$ and $y$ refer to the predictor vector and risk category, respectively. Then the prediction accuracy of LODO-CV is

$$\frac{1}{N}\sum_{k=1}^{N}\frac{1}{M}\sum_{m=1}^{M} I\left(y_m^k = \hat{f}^{-k}(x_m^k)\right)$$

where $I(x)$ is the indicator function. The prediction accuracy estimated by LODO-CV has two advantages. First, it maximized the size of the training set, i.e., 27 drugs, and thus reduced the power loss due to the validation. Second, it eliminated overfitting, as there are no drugs with observations in both training and test sets. The prediction of every left-out drug imitates the case that the model would predict a completely new drug.

### 3.3 Random Forest Model

The prediction of the drug-risk category is a binary classification problem in which, given a vector of predictors generated from the experiment, the predictive model will output the probability of the two risk categories (low or high). The risk category of one observation will be the risk with a higher probability. The probability of drug risk is the average of the observations belonging to that drug. Similarly, the risk category of one drug will be the risk with a higher probability. Note that this framework can accommodate any binary classifiers if they produce the probabilities of risk categories.

In this study, we utilized the random forest as the binary predictive model. Random forest is the ensemble of multiple decision trees that can capture the nonlinearity in a dataset (Breiman, 2001). A decision tree $T$ is a predictive model that assigns each observation to a certain class based on split roles defined on the predictor space. Formally, suppose there are $p$ predictors $\{X_1, X_2, \ldots, X_p\}$ in the dataset ($p = 16$ in the wedge dataset), and we split the predictor space into two regions $R_1$ and $R_2$ according to predictor $X_j$ and threshold $s$

$$R_1(j, s) = X|X_j \leq s$$

$$R_2(j, s) = X|X_j > s.$$

Then for any region $R_m$ with $N_m$ observations, let $\hat{p}_{mk}$ be the proportion of class $k$ in the region $R_m$

$$\hat{p}_{mk} = \frac{1}{N_m} \sum_{x_i \in R_m} I(y_i = k)$$

where $x_i$ and $y_i$ are the vector of predictors and risk category of observation $i$, respectively. $I(x)$ is the indicator function. The risk of any observation $x$ in the region $R_m$ is predicted as

$$\widehat{risk}(x) = argmax_k \hat{p}_{mk}$$

In each split that produces regions $R_1$ and $R_2$, we seek the predictor $X_j$ and threshold $s$ that solve the following optimization problem

$$min_{j,s} \left[ \sum_{x_i \in R_1(j,s)} L\big(\widehat{risk}(x_i), y_i\big) + \sum_{x_i \in R_2(j,s)} L\big(\widehat{risk}(x_i), y_i\big) \right]$$

where $L(x)$ is the loss function that can be misclassification error, Gini index, or cross-entropy (James et al, 2021). The splitting process is continued until some stopping rule is applied, usually the number of splits (tree depth) or the number of observations per region (leaf size).

To build a random forest from multiple decision trees, we generated $B$ ($B = 500$ in our implementation) bootstrap samples from the training data. We applied the previous splitting rule to build one decision tree per each bootstrap resample. Instead of searching all $p$ predictors, we randomly selected $\sqrt{p}$ predictors as candidates in each split to reduce the correlation among different decision trees. The final random forest $T_{b_1}^B$ contains $B$ decision trees, and the prediction of risk is the majority vote of all decision trees. Because the random forest is a combination of lowly correlated decision trees, it reduces the variance of prediction from a single decision tree and usually exhibits high prediction accuracy in real-world applications (Couronné, Probst, and Boulesteix, 2018).

## 4. RESULTS

### 4.1 Overall Model Performance

Based on the LODO-CV framework, we measured the model performance by prediction accuracy, the area under the receiver operating characteristic curve (AUROC), sensitivity, specificity, and precision. Suppose that there are $N$ observations in the dataset and $\hat{f}$ is the trained predictive model. Let $x_i$ and $y_i$ be the vector of predictors and scaler of the risk category for observation $i$. The prediction accuracy by observations is

$$acc_{obs} = \frac{1}{N}\sum_{i=1}^{N} I(\hat{f}(x_i) = y_i)$$

where $I(x)$ is the indicator function. The sensitivity, specificity, and precision are measurements of performance for one risk category. For any risk category $r$ and the other categories $\bar{r}$, define

$$sensitivity = \frac{\#\ of\ correct\ predictions\ in\ category\ r}{\#\ of\ observations\ in\ category\ r}$$

$$specificity = \frac{\text{\# of correct predictions in category } \bar{r}}{\text{\# of observations in category } \bar{r}}$$

$$precision = \frac{\text{\# of correct predictions in category } r}{\text{\# of observations with predicted category } r}$$

Note that the previous measurements can also be calculated on the drug level.

Figure 2, Table 1, and Table 2 summarize the prediction results of the random forest model on the wedge dataset. The random forest model achieves highly accurate prediction, with a binary prediction accuracy of 0.8750 by observations and 0.9286 by drugs. The AUROC of binary classification is 0.9338 by observations and 0.9880 by drugs, indicating a very strong binary classification capacity. In addition, the sensitivity, specificity, and recall of low and high risk are mostly greater than 0.8, both measured by observations and drugs. In summary, the random forest model can accurately predict the low and high TdP drug risks in the wedge dataset. We note that averaging the probabilities across observations improves the binary classification at the drug level. This demonstrates the benefits of repeated experimental design in preclinical drug proarrhythmic assessment (Keselman, Algina, and Kowalchuk, 2001; Wang, Xiao, and Xu, 2018; Xiao, Wang, and Xu, 2019).

## 4.2 Model Uncertainty Measurement

The previous result can be understood as a point estimate of the model performance. Beyond that, we conducted a resampling study by stratified bootstrap to further explore the uncertainty in the model prediction (Murphy, 2012). Suppose that $Z^{(k)} = \left(z_1^{(k)}, z_2^{(k)}, \ldots, z_M^{(k)}\right)$ are the observations of drug $k$, where $k = 1, 2, \ldots, 28$, and $M = 4$ in the wedge dataset. Let $Z^{(k)*} = \left(z_1^{(k)*}, z_2^{(k)*}, \ldots, z_M^{(k)*}\right)$ be a random sample drawn from $Z^{(k)}$ with replacement. Then the stratified bootstrap sample $Z^*$ is defined as the union of $Z^{(k)*}$, where $k = 1, 2, \ldots, 28$. We generated 1000 stratified bootstrap samples $Z^*$ and repeated the previous model training and evaluation process on each $Z^*$ to obtain empirical distributions of prediction accuracy, AUROC, sensitivity, specificity, and precision. The 95% confidence interval, mean, and standard deviation of five performance measurements are constructed based on their empirical distributions in Table 1 and Table 2. Compared with regular bootstrap, stratified bootstrap guarantees that the structure of repeated measurements for each drug was preserved in the resampling process. Table 1 and Table 2 show that the overall prediction accuracy, AUROC, sensitivity, specificity, and precision are stable in the stratified bootstrap process. The narrow confidence intervals indicate that the predictive model is consistent across different measurements. For each drug, we also calculated the 95% confidence interval for the probability of being low or high risks by averaging the probability of four observations in each drug (Supplementary Table S3).

### 4.3 Predictor importance

We conducted a permutation predictor importance analysis to explore the effects of different predictors on the random forest model. The permutation predictor importance is defined as the decrease in the model prediction accuracy when a single predictor value is randomly shuffled (Breiman, 2001). Because the random shuffling broke the relationship between predictors and target variables, the decrease of the prediction accuracy indicates the model's dependency on that predictor.

Suppose that there are $p$ predictors in the dataset $D$. We trained a model $\hat{f}$ on dataset $D$ and calculated the prediction accuracy $acc$ by observations using LODO-CV. We then randomly shuffled predictor $j$ to generate a permuted dataset $\widetilde{D}_j$ and calculated the permuted prediction accuracy $acc_j$ by applying model $\hat{f}$ to $\widetilde{D}_j$. The permutation importance of predictor $j$ $imp_j$ is calculated as

$$imp_j = acc - acc_j.$$

The same process was repeated 100 times, and the distributions and means of the permutation predictor importance for each predictor are shown in Figure 3. The ratio of QT interval to QS interval and ratio of the interval between the peak and the end of T wave (TPE) to QT interval, two critical endpoints in the experiment of drug risk assessment (Blinova et al, 2018; Liu et al, 2021), are the two most important predictors in the wedge dataset, respectively. We observed that the difference of predictor importance was moderate among predictors, possibly because all predictors in the dataset were selected by domain experts before the experiment.

### 4.4 Sensitivity Analysis and Visualization

As the prediction accuracy is below 100% by both observations and drugs, we examined which drugs the model predicted most inaccurately. There are two drugs with zero or half correct predictions among four observations in the wedge dataset (Table 3). We referred to these drugs as potential outlier drugs due to their low accuracy. Sensitivity analysis was conducted to evaluate the impact of potential outlier drugs on model performance. Specifically, we removed potential outlier drugs from the original datasets, repeated the modeling, and estimated the model performance following previous sections. As a result, the model obtained improved performance without potential outlier drugs (Table 4).

Furthermore, we conducted dimension reduction by applying t-distributed stochastic neighbor embedding (t-SNE) on the wedge dataset (van der Maaten and Hinton, 2008). Intuitively, t-SNE compresses the dataset from the original high-dimension space (16 in the wedge dataset) to the two-dimensional space and simultaneously preserves the similarity among observations. Observations similar to each other in the original high-dimensional space will be closed in the two-dimensional space. Figure 4 shows the wedge datasets in the t-SNE two-dimensional space. Ideally, observations from different risk categories will form distinctive clusters. The more separable observations are, the more accurately models can predict. In the wedge dataset, observations produced relatively distinctive clusters, and thus the model trained on it would predict accurately. This is consistent with the high model performance in Table 1 and Table 2. We also found that the two potential outlier drugs are mostly located at the boundary of two clusters or scattered between different clusters. Such structure lowered the separability; therefore, it was more difficult for predictive models to accurately assign risk categories.

**4.5 Control Analysis**

Control analysis is a common practice in experiments of TdP risk assessment (Gintant, 2011; Rampe and Brown, 2013)[20,21]. Motivated by that, we conducted a control analysis to further validate the model prediction. In the control analysis, the model will predict the test drug, positive control, and negative control simultaneously. We will only trust the model that can correctly predict both controls. If the model prediction of controls was incorrect, then its prediction on test drugs would be considered unreliable. The commonly used positive controls are high-risk drugs with a clearly understood mechanism, while negative control is a low-risk drug or placebo (treated as low risk). Control analysis provides another perspective of model validation. In our study, the high-risk drug dofetilide was selected as a positive control due to its similar role in the experiments. Mexiletine and placebo were included in the wedge dataset and thus selected as the negative control. It should be noted that these two drugs are examples of positive and negative drugs. We could also select other drugs with clearly understood risk mechanisms.

LODO-CV was applied to evaluate the model performance on 27 noncontrol drugs, positive control (dofetilide), and negative control (mexiletine and placebo) (Figure 5). Table 5 and Table 6 show the result of control analysis under the random forest model. We observed high prediction accuracy when the model correctly predicted positive control (dofetilide) and negative control (mexiletine). In the wedge dataset, the prediction accuracy by observations on the 27 noncontrol drugs is 0.9327, conditioning on the correct prediction of both positive and negative control. We also noted that the model predicted positive control (dofetilide) very accurately, with zero error in the wedge dataset. However, if we selected placebo as the negative control, only 15.74% of placebos were correctly identified as low risk in the wedge dataset. We suspected that the placebo might be different from other low-risk drugs in terms of electrophysical signals obtained from experiments.

## 5. CONCLUSION

In this study, we proposed a comprehensive machine learning framework to predict drug-induced TdP risk from preclinical protocols. The random forest model was trained on the isolated arterially perfused rabbit ventricular wedge preparation dataset. The unbiased estimate of model performance was measured by LODO-CV. The uncertainty of model performance was evaluated by stratified bootstrap. Our proposed method provided interpretability consistent with domain knowledge through permutation prediction importance. Sensitivity analysis identified potential outlier drugs, which were also described in the literature. We conducted control analysis and further validated the model performance. Overall, our statistical models exhibited high accuracy in predicting drug-induced TdP risks, with 0.8750 by observations and 0.9286 by drugs.

The potential outlier drugs identified by our models are consistent with domain knowledge and thus suggest the limitation of current experimental techniques. (Crumb, et al 2016; Johannesen, et al, 2016). For example, clarithromycin and ranolazine have significantly lower prediction accuracy than other drugs with a clearly understood mechanism. This result suggests the potential application of our modeling strategy to validate new experimental techniques. One supplement of the current study is to validate the model performance by cross-dataset validation (Hastie, Tibshirani, and Friedman, 2009). The predictive model trained on one dataset would be evaluated on another dataset generated by the same experimental protocol. Such evaluation will provide a more realistic estimation of model performance. Another potential extension is to utilize novel experimental techniques beyond electrophysiological signals (Xi and Li, 2021a,b). Single-cell RNA-sequencing can be applied to evaluate and compare the transcriptomes of individual cells before and after drug treatment. The gene expression dynamic associated with drug-induced TdP risk will potentially improve the assessment of such risks.

## ACKNOWLEDGMENTS

We would like to thank Dr. Gan-Xin Yan for his permission to use the wedge data from his lab. We are grateful for the valuable discussions with Drs. Qianyu Dang, Christine Garnett, Donglin Guo, Yu-yi Hsu, Lars Johannesen, Jose Vicente Ruiz, and Wendy Wu. Finally, we want to thank Drs. Yi Tsong and Atiar Rahman for their support for this project.

## REFERENCES

Blinova, K., Dang, Q., Millard, D., Smith, G., Pierson, J., Guo, L., Brock, M., Lu, H.R., Kraushaar, U., Zeng, H., Shi, H., Zhang, X., Sawada, K., Osada, T., Kanda, Y., Sekino, Y., Pang, L., Feaster, T.K., Kettenhofen, R., Stockbridge, N., Strauss, D.G., Gintant, G. (2018), "International Multisite Study of Human-Induced Pluripotent Stem Cell-Derived Cardiomyocytes for Drug Proarrhythmic Potential Assessment," *Cell Reports*, 24, 3582-3592. DOI:10.1016/j.celrep.2018.08.079.

Breiman, L. (1996), "Bagging predictors," *Machine Learning*, 24, 123–140. DOI: 10.1007/BF00058655.

Breiman, L. (2001), "Random Forests," *Machine Learning*, 45, 5–32. DOI: 10.1023/A:1010933404324.

Couronné, R., Probst, P., and Boulesteix, A.-L. (2018), "Random forest versus logistic regression: a large-scale benchmark experiment," *BMC Bioinformatics*,19, DOI: 10.1186/s12859-018-2264-5.

Crumb, W. J., Jr, Vicente, J., Johannesen, L., and Strauss, D. G. (2016), "An evaluation of 30 clinical drugs against the comprehensive in vitro proarrhythmia assay (CiPA) proposed ion channel panel," *Journal of Pharmacological and Toxicological Methods*, 81, 251–262. DOI: 10.1016/j.vascn.2016.03.009.

Fermini, B., Hancox, J.C., Abi-Gerges, N., Bridgland-Taylor, M., Chaudhary, K.W., Colatsky, T., Correll, K., Crumb, W., Damiano, B., Erdemli, G., Gintant, G., Imredy, J., Koerner, J., Kramer, J.,

Levesque, P., Li, Z., Lindqvist, A., Obejero-Paz, C.A., Rampe, D., Sawada, K., Strauss, D.G., and Vandenberg, J.I.. (2016), "A new perspective in the field of cardiac safety testing through the comprehensive in vitro proarrhythmia assay paradigm," *Journal of Biomolecular Screening*, 21, 1–11. DOI: 10.1177/1087057115594589.

Gintant, G. (2011), "An evaluation of hERG current assay performance: Translating preclinical safety studies to clinical QT prolongation," *Pharmacology and Therapeutics*, 129, 109–119. DOI: 10.1016/j.pharmthera.2010.08.008.

Hastie, T., Tibshirani, R., and Friedman, J. (2009), *The Elements of Statistical Learning: Data Mining, Inference, and Prediction* (2nd ed), New York, NY:Springer Science & Business Media.

James, G., Witten, D., Hastie, T. & Tibshirani, R. *An introduction to statistical learning*. (Springer, 2021).

Johannesen, L., Vicente, J., Mason, J. W., Erato, C., Sanabria, C., Waite-Labott, K., Hong, M., Lin, J., Guo, P., Mutlib, A., Wang, J., Crumb, W. J., Blinova, K., Chan, D., Stohlman, J., Florian, J., Ugander, M., Stockbridge, N., and Strauss, D.G. (2016), "Late sodium current block for drug-induced long QT syndrome: Results from a prospective clinical trial," *Clinical Pharmacology and Therapeutics*, 99, 214–223. DOI: 10.1002/cpt.205.

Keselman, H. J., Algina, J., and Kowalchuk, R. K. (2001), "The analysis of repeated measures designs: a review," *British Journal of Mathematical and Statistical Psychology*, 54, 1–20. DOI: 10.1348/000711001159357.

Li, Z., Ridder, B. J., Han, X., Wu, W. W., Sheng, J., Tran, P. N., Wu, M., Randolph, A., Johnstone, R. H., Mirams, G. R., Kuryshev, Y., Kramer, J., Wu, C., Crumb, W. J. Jr, and Strauss, D. G. (2019), "Assessment of an in silico mechanistic model for proarrhythmia risk prediction under the CiPA initiative," *Clinical Pharmacology and Therapeutics*, 105, 466–475. DOI: 10.1002/cpt.1184.

Liu, T., Brown, B.S., Wu, Y., Antzelevitch, C., Kowey, P. R., and Yan, G.X. (2006), "Blinded validation of the isolated arterially perfused rabbit ventricular wedge in preclinical assessment of drug-induced proarrhythmias," *Heart Rhythm*, 3, 948–956. DOI: 10.1016/j.hrthm.2006.04.021.

Liu, T., Traebert, M., Ju, H., Suter, W., Guo, D., Hoffmann, P., Kowey, P. R., Yan, G. X. (2013) "Differentiating electrophysiological effects and cardiac safety of drugs based on in vitro electrocardiogram: A blinded validation," *Journal of Pharmacological and Toxicological Methods*, 68, e23. DOI: 10.1016/j.hrthm.2012.06.030.

Liu, T., Liu, J., Lu, H.R., Li, H., Gallacher, D.J., Chaudhary, K., Wang, Y., and Yan, G.X. (2021), "Utility of Normalized TdP Score System in drug proarrhythmic potential assessment: A blinded in vitro study of CiPA drugs," *Clinical Pharmacology and Therapeutics*, 109, 1606–1617. DOI: 10.1002/cpt.2133.

Murphy, K. P. (2012), *Machine Learning: A Probabilistic Perspective*. Cambridge, MA:MIT Press.

Rampe, D. and Brown, A. M. (2013), "A history of the role of the hERG channel in cardiac risk assessment," *Journal of Pharmacological and Toxicological Methods*, 68, 13–22. DOI: 10.1016/j.vascn.2013.03.005.

Sager, P. T., Gintant, G., Turner, J. R., Pettit, S., and Stockbridge, N. (2014), "Rechanneling the cardiac proarrhythmia safety paradigm: a meeting report from the Cardiac Safety Research Consortium." *American Heart Journal*, 167, 292–300. DOI: 10.1016/j.ahj.2013.11.004.

Stockbridge, N., Morganroth, J., Shah, R.R., and Garnett, C. (2013), "Dealing with global safety issues : was the response to QT-liability of non-cardiac drugs well coordinated?" *Drug Safety*, 36,167-182. DOI: 10.1007/s40264-013-0016-z.

van der Maaten, L. and Hinton, G. (2008), "Visualizing Data using t-SNE," *Journal of Machine Learning Research*, 9, 2579–2605.

Wang, L., Xiao, Q., and Xu, H. (2018), "Optimal maximin $L_1$-distance Latin hypercube designs based on good lattice point designs," *Annals of Statistics*, 46, 3741-3766. DOI: 10.1214/17-AOS1674.

Xi, N. M., Hsu, Y.-Y., Dang, Q., and Huang, D. P. (2021), "Statistical Learning in Preclinical Drug Proarrhythmic Assessment," *arXiv [stat.AP]*.

Xi, N. M. & Li, J. J. (2021a), "Benchmarking Computational Doublet-Detection Methods for Single-Cell RNA Sequencing Data," *Cell Systems*, 12, 176-194.e6. DOI: 10.1016/j.cels.2020.11.008.

——— (2021b), "Protocol for executing and benchmarking eight computational doublet-detection methods in single-cell RNA sequencing data analysis," STAR Protocols, 2, 100699. DOI: 10.1016/j.xpro.2021.100699.

Xiao, Q., Wang, L. & Xu, H. (2019), Application of kriging models for a drug combination experiment on lung cancer. *Statistics in Medicine,* 38, 236–246. DOI: 10.1002/sim.7971.

**Figures and Tables**

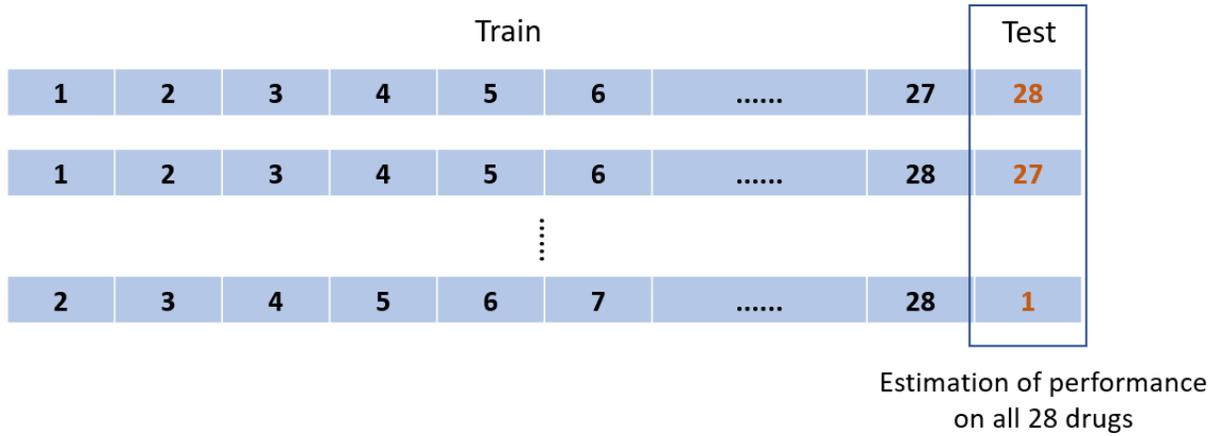

**Figure 1.** The schematic diagram of leave-one-drug-out cross-validation. In each iteration, the predictive model was trained on 27 drugs and predicted on the one left-out drug. The process was repeated until each drug was predicted, and the model performance was calculated by combining the prediction result of each drug.

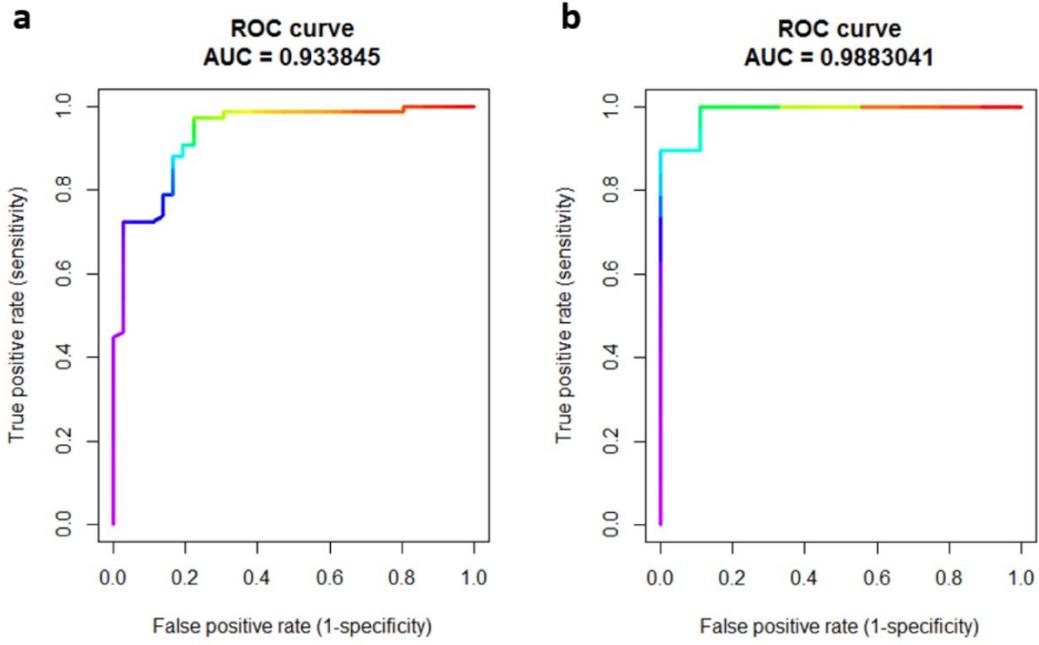

**Figure 2.** The ROC curve and AUROC of random forest model on the wedge dataset. a, ROC curve and AUROC measured by observations; b, ROC curve and AUROC measured by drugs.

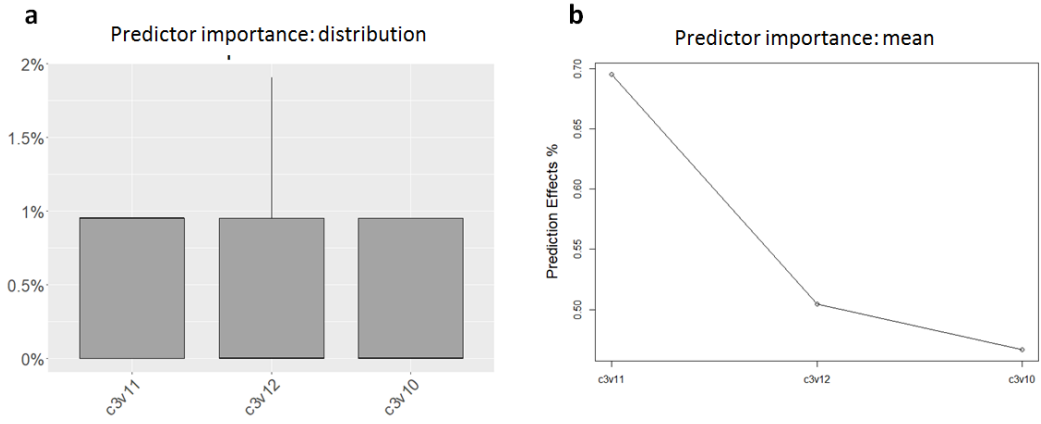

**Figure 3.** The distributions and averages of permutation predictor importance calculated under the random forest model by observation: a The distribution of the top three variables; b. The average of the top three variables.

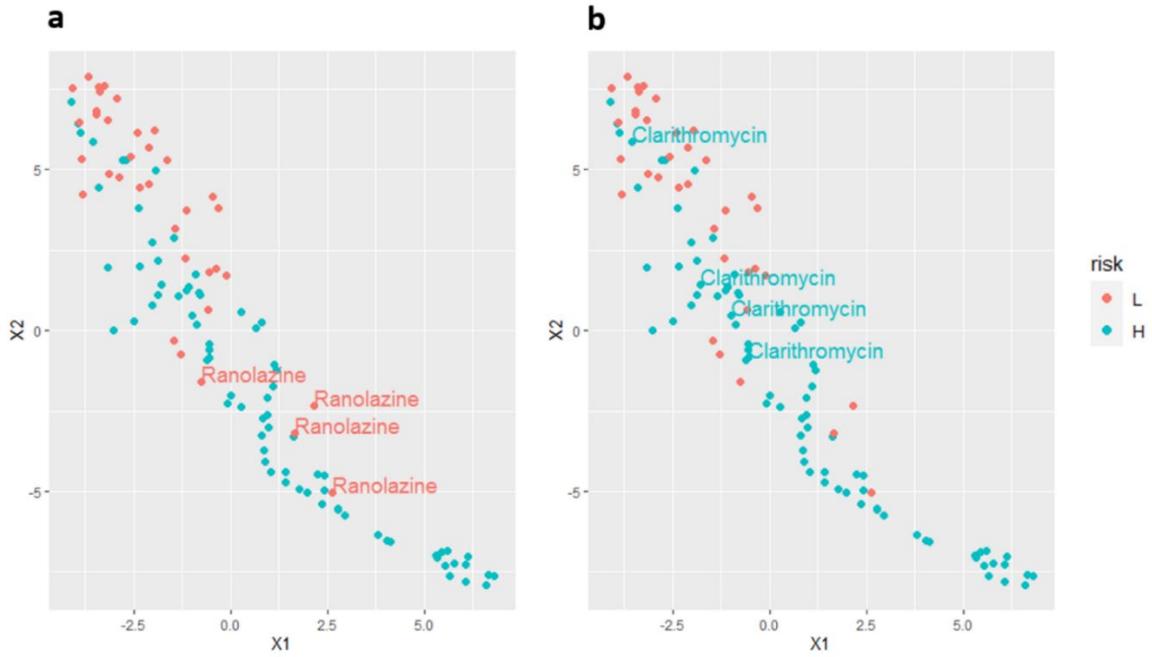

**Figure 4.** The two-dimensional t-SNE plot of the wedge dataset and two potential outlier drugs: a, Ranolazine; b, Clarithromycin.

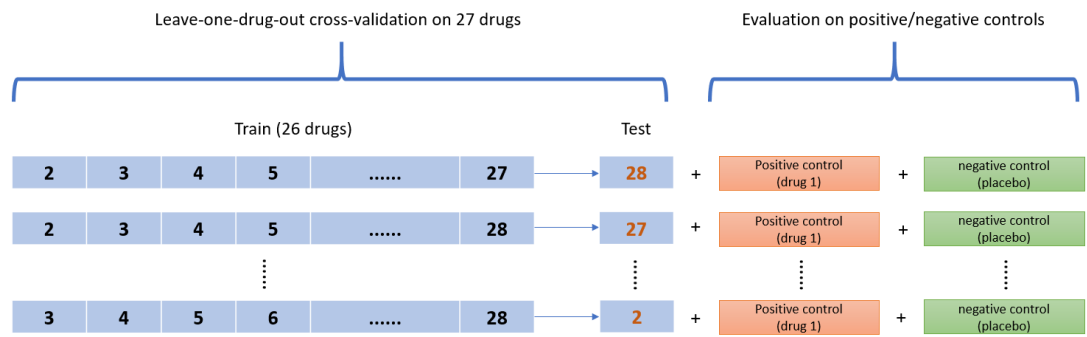

**Figure 5.** The schematic diagram of control analysis. Leave-one-drug-out cross-validation was applied to estimate the prediction accuracy of 27 noncontrol drugs, positive control, and negative control (placebo).

**Table 1.** The performance and uncertainty of the random forest model on the wedge dataset. All measurements were calculated by observations.

|  | Overall Accuracy | AUROC | High risk | | | Low risk | | |
|---|---|---|---|---|---|---|---|---|
|  |  |  | Sensitivity | Specificity | Precision | Sensitivity | Specificity | Precision |
| Point estimate | 0.8750 | 0.9338 | 0.9211 | 0.7778 | 0.8974 | 0.7778 | 0.9211 | 0.8235 |
| 95% CI | (0.8571, 0.9107) | (0.9177, 0.9521) | (0.8947, 0.9473) | (0.7778, 0.8333) | (0.8947, 0.9210) | (0.7778, 0.8333) | (0.8947, 0.9473) | (0.7778, 0.8823) |
| Mean (bootstrap) | 0.8811 | 0.9336 | 0.9223 | 0.7939 | 0.9043 | 0.7939 | 0.9223 | 0.8301 |
| Sd (bootstrap) | 0.0129 | 0.0090 | 0.0174 | 0.0168 | 0.0072 | 0.0168 | 0.0174 | 0.0324 |

**Table 2.** The performance and uncertainty of the random forest model on the wedge dataset. All measurements were calculated by drugs.

|  | Overall Accuracy | High risk | | | Low risk | | |
|---|---|---|---|---|---|---|---|
|  |  | Sensitivity | Specificity | Precision | Sensitivity | Specificity | Precision |
| Point estimate | 0.9286 | 0.9474 | 0.8889 | 0.9474 | 0.8889 | 0.9474 | 0.8889 |
| 95% CI | (0.8571, 1) | (0.8421, 1) | (0.8889, 1) | (0.9411, 1) | (0.8889, 1) | (0.8421, 1) | (0.7272, 1) |
| Mean (bootstrap) | 0.9250 | 0.9210 | 0.9333 | 0.9674 | 0.9333 | 0.9210 | 0.8552 |
| Sd (bootstrap) | 0.0384 | 0.0481 | 0.0569 | 0.0278 | 0.0481 | 0.0569 | 0.0815 |

**Table 3.** Drugs with the lowest prediction accuracy on the wedge dataset. The correct rate is the proportion of correct prediction among observations per drug and was calculated based on the random forest model.

| Dataset | Drug | Correct rate | True risk | Predicted risk |
|---|---|---|---|---|
| Wedge | Clarithromycin | 0.00 | High | Low |
|  | Ranolazine | 0.50 | Low | High |

**Table 4.** The model performance before and after dropping potential outlier drugs.

| Dataset | Model |  | With potential outlier drugs | Without potential outlier drugs |
|---|---|---|---|---|
| Wedge | Random forest | By observations | 0. 8750 | 0.9808 |
|  |  | By drugs | 0.9286 | 1.0000 |
|  |  | AUROC (by observations) | 0.9371 | 0.9941 |

**Table 5.** Control analysis results on the wedge dataset with placebo as the negative control.

| | Positive control | Negative control | Original | 27 Test drugs (without control) | Positive control acc | Negative control acc | Test drugs (condition on correct positive control) | Test drugs (condition on correct negative control) | Test drugs (condition on both correct positive and negative control) |
|---|---|---|---|---|---|---|---|---|---|
| By obs | Dofetilide | Placebo | 0.8750 | 0.8750 | 1.0000 | 0.1574 | 0.9352 | 0.8824 | 0.8824 |
| By drugs | Dofetilide | Placebo | 0.9286 | 0.9259 | - | - | 0.9259 | 0.8333 | 0.8333 |

**Table 6.** Control analysis results on the wedge dataset with mexiletine as the negative control.

| | Positive control | Negative control | Original | 26 Test drugs (without control) | Positive control acc | Negative control acc | Test drugs (condition on correct positive control) | Test drugs (condition on correct negative control) | Test drugs (condition on both correct positive and negative control) |
|---|---|---|---|---|---|---|---|---|---|
| By obs | Dofetilide | Mexiletine | 0.8750 | 0.8750 | 1.0000 | 1.0000 | 0.9327 | 0.9327 | 0.9327 |
| By drugs | Dofetilide | Mexiletine | 0.9286 | 0.9230 | - | - | 0.9321 | 0.9321 | 0.9321 |

# SUPPLEMENTARY

Table S1. The 28 drugs in the wedge dataset and used in this study.

| Low TdP Risk (9) | Intermediate TdP Risk (11) | High TdP Risk (8) |
|---|---|---|
| Diltiazem, Loratadine Ranolazine, Metoprolol Mexiletine, Nifedipine Nitrendpine, Tamoxifen Verapamil | Astemizole, Chlorpromazine Cisapride, Clarithromycin Clozapine, Domperidone, Droperidol, Ondansetron Pimozide, Risperidone, Terfenadine | Azimilide, Bepridil, Disopyramide, Dofetilide, Ibutilide, Quinidine, Sotalol, Vandetanib |

**Table S2.** The 16 predictors in the wedge dataset. Except for extra notes, all predictors were obtained from an electrocardiogram at a 2000 pace rate.

| Predictor | Description | Type |
|---|---|---|
| c3v1 | TdP score proposed in the original study (Couronné, Probst, and Boulesteix, 2018) | Continuous |
| c3v2 | Interval between the J point and the start of T wave (JT) | Continuous |
| c3v3 | Interval between the J point and the peak of T wave (JTP) | Continuous |
| c3v4 | Ratio of the JTP interval at concentration $\geq 0$ to concentration $= 0$ | Continuous |
| c3v5 | QS interval, measured at pace rate 1000 | Continuous |
| c3v6 | QS interval, measured at pace rate 2000 | Continuous |
| c3v7 | QS interval, measured at pace rate 500 | Continuous |
| c3v8 | Ratio of the QS interval at concentration $\geq 0$ to concentration $= 0$, measured at pace rate 500 | Continuous |
| c3v9 | Interval between Q point and the end of T wave (QTE) | Continuous |
| c3v10 | Ratio of the QTE interval at concentration $\geq 0$ to concentration $= 0$ | Continuous |
| c3v11 | Ratio of QT interval to QS interval | Continuous |
| c3v12 | Ratio of the interval between the peak and the end of T wave (TPE) to QT interval | Continuous |
| c3v13 | Ratio of TPE interval to QT interval | Continuous |
| c3v14 | TPE Interval | Continuous |
| c3v15 | Ratio of TPE interval at concentration $\geq 0$ to concentration $= 0$ | Continuous |
| c3v16 | Score as a function of early afterdepolarizations (EADs) (Couronné, Probst, and Boulesteix, 2018) | Continuous |

**Table S3.** The point estimation of being low or high risk, and their 95% confidence interval for each drug. The confidence interval was constructed by stratified bootstrap.

| Drug | Risk | Probability of low risk | | | Probability of high risk | | |
|---|---|---|---|---|---|---|---|
| | | Point estimate | 95% CI lower bound | 95% CI upper bound | Point estimate | 95% CI lower bound | 95% CI upper bound |
| Astemizole | H | 0.2440 | 0.1495 | 0.3680 | 0.7560 | 0.6340 | 0.8505 |
| Azimilide | H | 0.0060 | 0.0025 | 0.0215 | 0.9940 | 0.9785 | 0.9975 |
| Bepridil | H | 0.0125 | 0.0025 | 0.0150 | 0.9875 | 0.9850 | 0.9975 |
| Chlorpromazine | H | 0.4295 | 0.3155 | 0.6070 | 0.5705 | 0.3930 | 0.6845 |
| Cisapride | H | 0.1980 | 0.1650 | 0.2545 | 0.8020 | 0.7455 | 0.8350 |
| Clarithromycin | H | 0.0330 | 0.0240 | 0.0705 | 0.9670 | 0.9295 | 0.9760 |
| Clozapine | H | 0.0235 | 0.0185 | 0.0640 | 0.9765 | 0.9360 | 0.9815 |
| Diltiazem | L | 0.9345 | 0.8620 | 0.9485 | 0.0655 | 0.0515 | 0.1380 |
| Disopyramide | H | 0.0080 | 0.0005 | 0.0255 | 0.9920 | 0.9745 | 0.9995 |
| Dofetilide | H | 0.0000 | 0.0000 | 0.0030 | 1.0000 | 0.9970 | 1.0000 |
| Domperidone | H | 0.0180 | 0.0070 | 0.0420 | 0.9820 | 0.9580 | 0.9930 |
| Droperidol | H | 0.0020 | 0.0010 | 0.0140 | 0.9980 | 0.9860 | 0.9990 |
| Ibutilide | H | 0.0010 | 0.0000 | 0.0020 | 0.9990 | 0.9980 | 1.0000 |
| Loratadine | L | 0.5190 | 0.4530 | 0.6065 | 0.4810 | 0.3935 | 0.5470 |
| Metoprolol | L | 0.8870 | 0.8000 | 0.9155 | 0.1130 | 0.0845 | 0.2000 |
| Mexiletine | L | 0.9790 | 0.9560 | 0.9890 | 0.0210 | 0.0110 | 0.0440 |
| Nifedipine | L | 0.9790 | 0.9465 | 0.9920 | 0.0210 | 0.0080 | 0.0535 |
| Nitrendpine | L | 0.6980 | 0.6290 | 0.8035 | 0.3020 | 0.1965 | 0.3710 |
| Ondansetron | H | 0.0280 | 0.0155 | 0.0580 | 0.9720 | 0.9420 | 0.9845 |
| Pimozide | H | 0.1605 | 0.1055 | 0.2340 | 0.8395 | 0.7660 | 0.8945 |
| Quinidine | H | 0.0045 | 0.0005 | 0.0195 | 0.9955 | 0.9805 | 0.9995 |
| Ranolazine | L | 0.3275 | 0.2695 | 0.4185 | 0.6725 | 0.5815 | 0.7305 |
| Risperidone | H | 0.0145 | 0.0080 | 0.0395 | 0.9855 | 0.9605 | 0.9920 |
| Sotalol | H | 0.0060 | 0.0000 | 0.0135 | 0.9940 | 0.9865 | 1.0000 |
| Tamoxifen | L | 0.6645 | 0.5190 | 0.7650 | 0.3355 | 0.2350 | 0.4090 |
| Terfenadine | H | 0.5165 | 0.3940 | 0.6070 | 0.4835 | 0.3930 | 0.6060 |
| Vandetanib | H | 0.0055 | 0.0025 | 0.0180 | 0.9945 | 0.9820 | 0.9975 |
| Verapamil | L | 0.8750 | 0.7735 | 0.8905 | 0.1250 | 0.1095 | 0.2265 |